\documentclass[a4paper,12pt ]{article}
\usepackage {amssymb,amsmath,latexsym,enumerate,graphicx}
\usepackage{indentfirst}
\usepackage{titlesec}
\usepackage {subfigure,epsfig}
\usepackage{caption2}
\numberwithin{equation}{section}
\usepackage {cite}
\begin{document}
\title{Power expansions for the self-similar solutions
of the modified Savada-Kotera  equation. }
\author{
Olga Yu. Efimova \and Nikolai A. Kudryashov }
\date{} \maketitle

\begin{abstract}
The fourth-order ordinary differential equation, defining new
transcendents, is studied.  The self-similar solutions of the
Kaup-Kupershmidt and Savada-Kotera equations are shown to be found
taking its solutions into account. Equation studied belongs to the
class of fourth-order analogues of the Painlev\'{e} equations. All
the power and non-power asymptotic forms and expansions near points
$z=0$, $z=\infty$ and near arbitrary point $z=z_0$ are found by
means of power geometry methods. The exponential additions to the
solutions of studied equation are also determined.
\end{abstract}

\section{Introduction.}
We consider the fourth-order ordinary equation
\cite{Hone01,Mugan01,Mugan02,Kudryashov01,Cosgrove01,Kudryashov07,Kudryashov10}
\begin{equation}\label{e:eq}
f(z,w)\stackrel{def}{=}w_{zzzz} - 5w^2w_{zz} +5 w_{z}w_{zz}- 5ww^2_z
+  w^5  - zw - \alpha =0
\end{equation}
which is one of the fourth-order analogues of the Painlev\'{e}
equations \cite{Mugan01,Mugan02,Kudryashov01,Cosgrove01,Kudryashov07,Kudryashov10}. It belongs to the
class of exactly solvable equations  and possesses B\"{a}cklund
transformations, Lax pair, rational and special solutions at some values
of parameter $\alpha$ \cite{Hone01,Mugan01,Mugan02,Kudryashov01,Cosgrove01,Kudryashov07,Kudryashov10}.

More than one century ago Painlev\'{e} and his collaborators
analyzed a certain class of the second-order nonlinear ordinary
differential equations. Their aim was to find all the second-order
canonical equations with the solutions without movable critical
points and to pick out those equations which have solutions defining
new special functions. As a result of investigations Painlev\'{e}
and his school discovered six second-order nonlinear ordinary
differential equations which solutions could not be expressed in
terms of known elementary or special functions. Nowadays they go
under the name of the Painlev\'{e} equations and their solutions are
called the Painlev\'{e} transcendents.

In the sixtieth of the twentieth century it was shown that the
Painlev\'{e} transcendents arose in the models describing physical
phenomena as frequently as many other special functions
\cite{Ablowitz01, Barouch03, Brezin04, De_Boer05, Ablowitz06,
Hall07, Chandrasecar08, Kudryashov01, Kudryashov02}. This fact
caused a significant interest to the studying of their properties
and set a problem to find other nonlinear equations defining new
transcendents. Equation \eqref{e:eq} belongs to the class of such
equations.

Let us also show that the self-similar solutions of the
Kaup-Kupershmidt equation   \cite{Caudrey01,Weiss01} and the
Savada-Kotera equation \cite{Sawada01,Weiss01} can be determined
by means of  ordinary differential equation \eqref{e:eq}.

The Kaup-Kupershmidt and the Savada-Kotera equations can be written as
\begin{equation}
u_t+u_{xxxxx}+10u u_{xxx}+10 \nu u_x u_{xx}+20 u^2 u_x=0
\end{equation}
where we consider the Kaup-Kupershmidt equation, if $\nu=1$, and
the Savada-Kotera  equation, if $\nu=5$ \cite{Parker01,Foursov01}.

Using the self-similar variables
\begin{equation}u(z)=\frac{1}{(5t)^{2/5}}g(z),\qquad z=\frac{x}{(5t)^{1/5}} \end{equation}
we obtain \begin{equation}E_1\equiv g_{zzzzz}+10 g
g_{zzz}+10g_zg_{zz}+20g^2g_z-2g-zg_z=0
\end{equation}
and
\begin{equation}E_2\equiv g_{zzzzz}+10 g g_{zzz}+25g_zg_{zz}+20g^2g_z-2g-zg_z=0
\end{equation}
for the Kaup-Kupershmidt and Savada-Kotera equations accordingly.

By means of the Miura transformation
\begin{equation}g=-(w_z+w^2)/2\end{equation}
we obtain that the Kaup-Kupershmidt equation in the self-similar variables can be
presented as
\begin{equation}E_1=\frac{1}{2}\left( \frac{d}{dz}+2w\right)\frac{d}{dz}f(z,w)=0 \end{equation}

Using the Miura transformation in the form
\begin{equation}g=w_z-w^2/2\end{equation}
we can also express the Savada-Kotera equation via equation \eqref{e:eq}
\begin{equation}E_2=\left(\frac{d}{dz}-w\right)\frac{d}{dz}f(z,w)=0 \end{equation}

So we have that the self-similar solutions of both the
Kaup-Kupershmidt and Savada-Kotera  equations can be determined by
means of  ordinary differential equation \eqref{e:eq}.

Painlev\'{e} equations determine the transcendental functions, so equation \eqref{e:eq}
is likely to do it too. However there is no exact proof of this statement. So it is
important to explore the asymptotic forms and expansions of this equation.

The aim of this paper is to calculate all asymptotic forms and
expansions to solutions of equation \eqref{e:eq}. To achieve it we
use the power geometry methods
\cite{Bruno01,Bruno02,Bruno03,Bruno08}.

The outline of this paper is as follows. In section 2 the general
properties of equation \eqref{e:eq} are discussed. In sections
3--8 asymptotic forms and expansions to the studied equation near
points $z=0$ and $z=\infty$ are given. It sections 9--12
exponential additions for these expansions are found. Finally,
section 13 is devoted to asymptotic forms and power expansions to
solutions of equation \eqref{e:eq} near arbitrary point $z=z_0$.

\section{General properties of equation \eqref{e:eq}.}
The monomials of equation \eqref{e:eq} are corresponded to points
$M_1=(-4,1)$, $M_2=(-2,3)$, $M_3=(-3,2)$, $M_4=(-2,3)$, $
M_5=(0,5)$, $M_6=(1,1)$ and $M_7=(0,0)$ (if $\alpha\ne0$).
Therefore, the carrier $S(f)$ of equation \eqref{e:eq} contains
points $Q_1=M_1$, $Q_2=M_5$, $Q_3=M_6$, $Q_4=M_8$ (if $\alpha\ne
0$), $Q_5=M_3$ and $Q_6=M_2=M_4$.

\begin{figure}[h] 
\center%
\epsfig{file=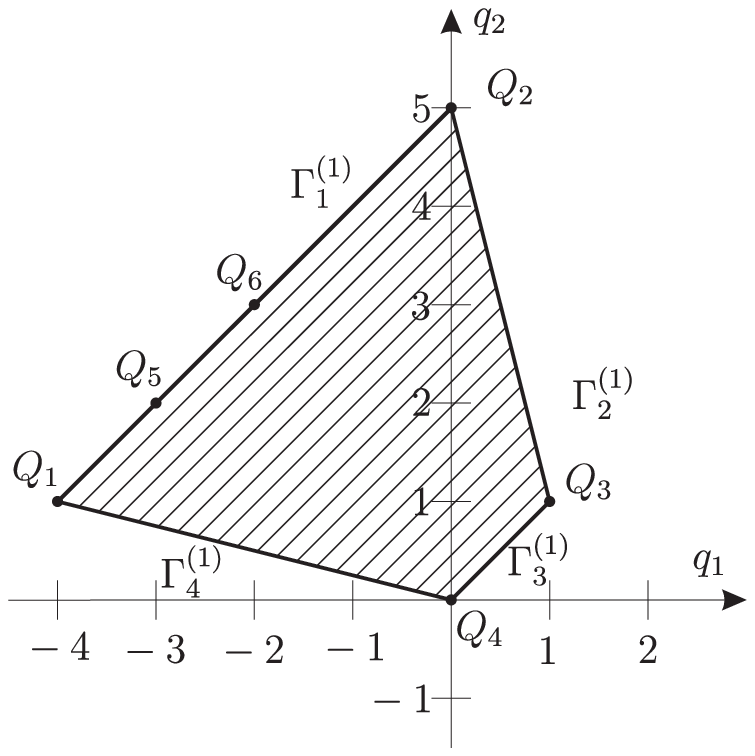,width=70mm} \caption{The carrier of equation
\eqref{e:eq} at $\alpha\ne0$}\label{fig:ff1}
\end{figure}

\begin{figure}[h] 
\center%
\epsfig{file=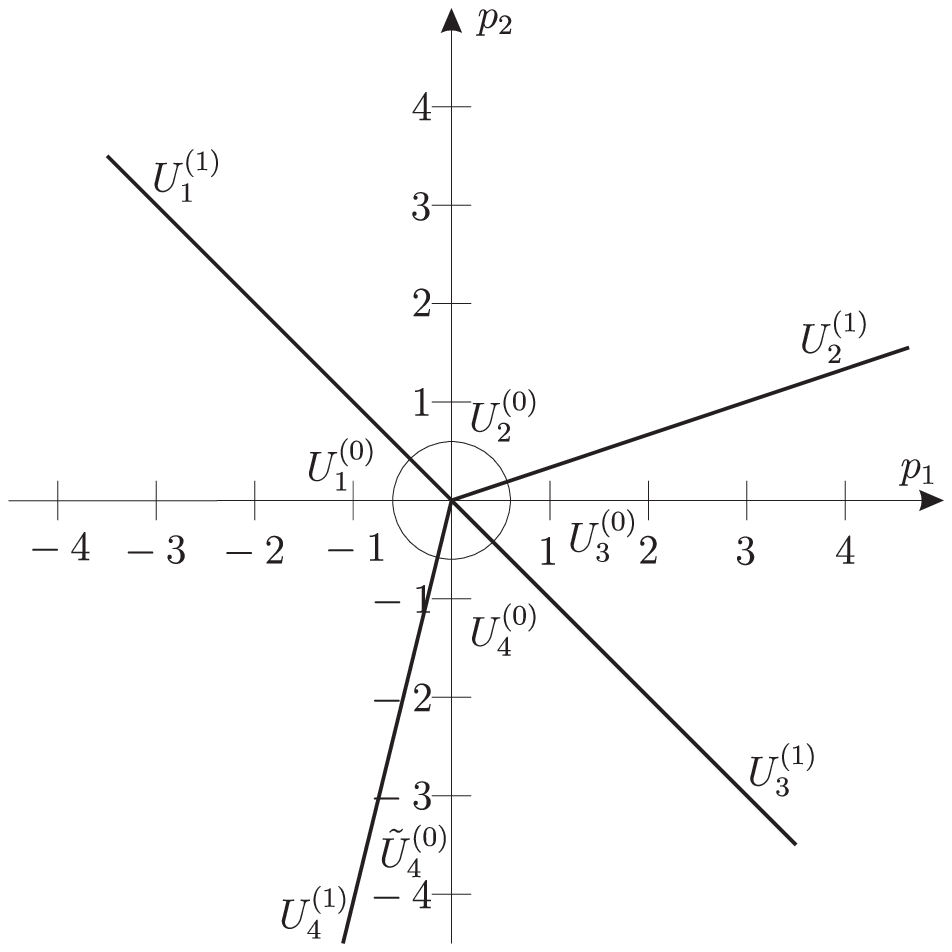,width=70mm} \caption{The normal cones of
equation \eqref{e:eq} at $\alpha\ne0$}\label{fig:ff2}
\end{figure}

In the case $\alpha\ne0$ the convex hull of the carrier of equation
\eqref{e:eq} is the quadrangle with four apexes
$\Gamma_j^{(0)}=Q_j\,(j=1,2,3,4)$ and four edges
$\Gamma_1^{(1)}=[Q_1,Q_2]$, $\Gamma_2^{(1)}=[Q_2,Q_3]$,
$\Gamma_3^{(1)}=[Q_3,Q_4]$, $\Gamma_4^{(1)}=[Q_1,Q_4]$. It is
presented at fig. \ref{fig:ff1}. The external normal vectors
$N_j\,(j=1,2,3,4)$ to edges $\Gamma_j^{(1)}\,(j=1,2,3,4)$ are
$N_1=(-1,1),\, N_2=(4,1),\, N_3=(1,-1),\, N_4=(-1,-4)$. They form
the normal cones $U_j^{(1)}$ of edges $\Gamma_j^{(1)}$.
\begin{equation}
\label{2.1}U_j^{(1)} =\{P=\lambda N_j,\,\,\, \lambda>0\},\,\,\,
j=1,2,3,4.
\end{equation}
The normal cones $U_j^{(0)}$ of apexes $\Gamma_j^{(0)}=Q_j\,\,
(j=1,\,2,\,3,\,4)$ are the angles between the edges that adjoin to
the apex. They  are given on fig. \ref{fig:ff2}.

\begin{figure}[h] 
\center%
\epsfig{file=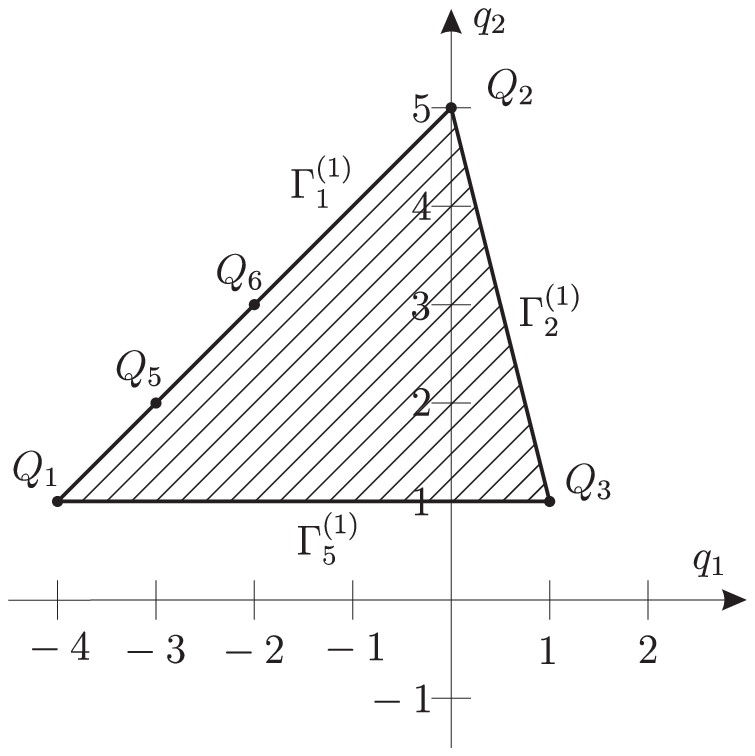,width=75mm} \caption{The carrier of equation
\eqref{e:eq} at $\alpha=0$}\label{fig:ff3}
\end{figure}

\begin{figure}[h] 
\center%
\epsfig{file=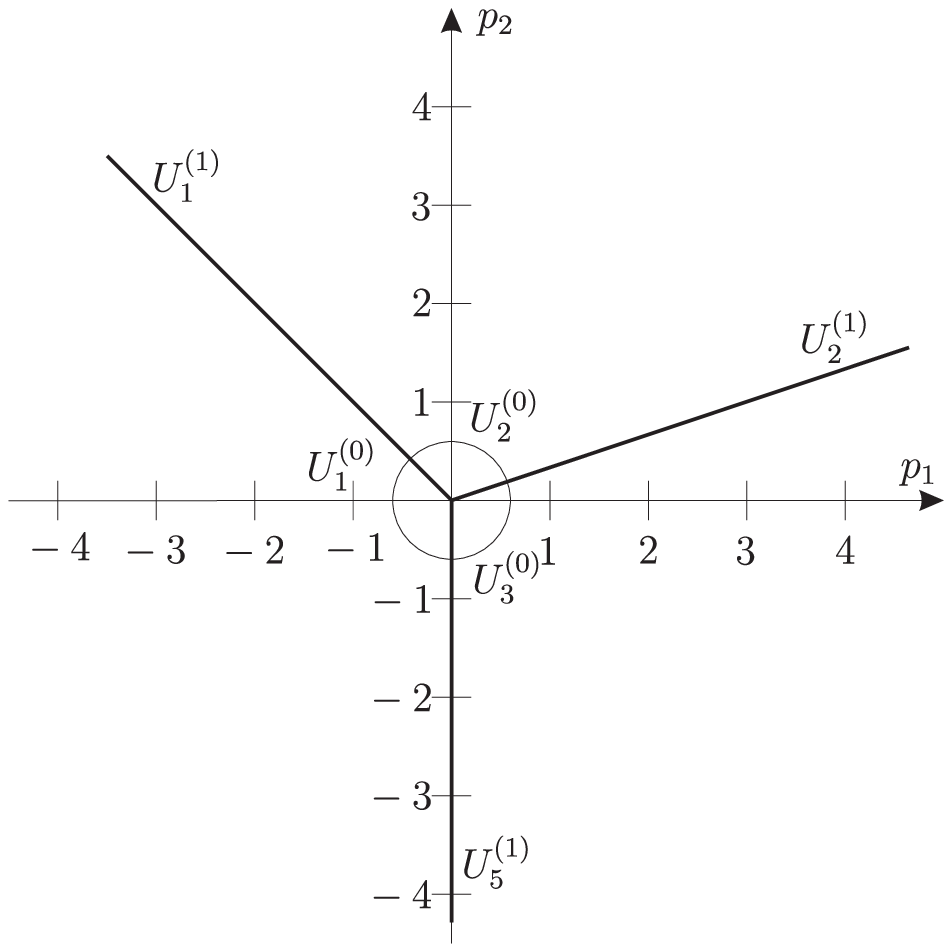,width=75mm} \caption{The normal cones of
equation \eqref{e:eq} at $\alpha=0$}\label{fig:ff4}
\end{figure}

If $\alpha=0$, the convex hull of the carrier of equation
\eqref{e:eq} is the triangle with apexes
$\Gamma_j^{(0)}=Q_j\,(j=1,2,3)$ and edges
$\Gamma_1^{(1)}=[Q_1,Q_2]$, $\Gamma_2^{(1)}=[Q_2,Q_3]$,
$\Gamma_5^{(1)}=[Q_3,Q_1]$ (fig. \ref{fig:ff3}). The normal cones
of equation \eqref{e:eq} in this case are presented at fig.
\eqref{fig:ff4}.

Lattice $\mathbb{Z}$, where the carrier of equation \eqref{e:eq}
lies, is generated by basis vectors $(-4,1)$ and $(1,1)$.

Examining the reduced equations corresponding to bounds
$\Gamma_j^{(d)}\,\, (d=0,1;\,\, j=1,\,2,\,3,\,4)$ we will obtain
the expansions of equation \eqref{e:eq}.

The reduced equations corresponding to apexes $\Gamma_2^{(0)}$,
$\Gamma_3^{(0)}$ and $\Gamma_4^{(0)}$
\begin{equation}
\begin{gathered}
\label{2.4}\hat{f}_2^{(0)}\stackrel{def}{=}w^5=0\\
\hat{f}_3^{(0)}\stackrel{def}{=} -zw=0\\
\hat{f}_4^{(0)}\stackrel{def}{=}-\alpha=0
\end{gathered}
\end{equation}
are the algebraic ones and hence they do not have non-trivial
power or non-power solutions.

\section{ Solutions, corresponding to apex $\Gamma_1^{(0)}$.}

Apex $\Gamma_1^{(0)}$ defines the following reduced equation
\begin{equation}
\label{3.1}\hat{f}_1^{(0)}\stackrel{def}{=} w_{zzzz}=0
\end{equation}
Let us find the reduced solutions
\begin{equation}
\label{3.2}w=c_r z^r,\,\,\, c_r\neq0
\end{equation}
for $\omega (1,r) \in U_1^{(0)}$. Since $p_1 < 0$ in the cone
$U_1^{(0)}$ then $\omega=-1$, $z\rightarrow 0$ and the expansions
will be the ascending power series of $z$. Substituting
\eqref{3.2} into $\hat{f}_1^{(0)}$ and cancelling the result by
$z^{r-4}$ we get the characteristic equation
\begin{equation}
\label{3.3}\chi(r)\stackrel{def}{=} r(r-1)(r-2)(r-3)=0
\end{equation}
with four roots $r_1=0,\,r_2=1,\,r_3=2,\,$ and $r_4=3$. Further we
shall examine them separately.

Vector $R=(1,0)$ that corresponds to the root $r_1=0$ being
multiplied by $\omega=-1$ belongs to the cone $U_1^{(0)}$. Thus we
obtain the family $\mathcal{F}_1^{(0)}\,1$ of power asymptotic
forms $w=c_0$ with the arbitrary constant $c_0 \neq 0$. The first
variation of equation \eqref{3.1}
\begin{equation}
\label{3.4}\frac{\delta\hat{f}_1^{(0)}}{\delta w} = \frac{d^4}{dz^4}
\end{equation}
gives the operator
\begin{equation}
\label{3.5}L(z)=\frac{d^4}{dz^4} \neq 0
\end{equation}
with the characteristic polynomial
\begin{equation}
\label{3.6}\nu(k) =z^{4-k} L(z) z^k =k(k-1)(k-2)(k-3)
\end{equation}
The equation $ \nu(k)=0 $ has four roots $k_1=0,\,\,\,
k_2=1,\,\,\, k_3=2,\,\,\,$и $k_4=3$. As $\omega=-1$ and $r=0$,
then the cone of the problem is $\mathcal{K}=\{k>0\}$. It contains
$k_2=1,\,k_3=2,\,k_3=3$, which are the critical numbers.

Thus the expansion of the solution corresponding to the reduced
equation \eqref{3.1} at $r=0$ takes the form
\begin{equation}
\label{3.10}w(z)= c_0 + c_1 z + c_2z^2 + c_3z^3 +
\sum^{\infty}_{k=4} c_k z^k
\end{equation}
where all the coefficients are constants, $c_0\neq
0,\,c_1,\,c_2,\,c_3$ are the arbitrary ones and $c_k \,\, (k\geq
4)$ are uniquely defined. Denote this family as $G_1^{(0)}\,1$.
Taking into account six terms, expansion \eqref{3.10} can be
presented in the form
\begin{equation*}\begin{gathered}
w(z)=c_{{0}}+c_{{1}}z+c_{{2}}{z}^{2}+c_{{3}}{z}^{3}+ \\+ \left(
\alpha-10\,c_{{1}}c_{{2}}-{c_{{0}}}
^{5}+10\,c_{{2}}{c_{{0}}}^{2}+5\,c_{{0}}{c_ {{1}}}^{2} \right)
\frac{{z}^{4}}{24}+\\
+ \left(c_{{0}}
-20\,{c_{{2}}}^{2}+40\,c_{{2}}c_{{0}}c_{{1}}+30\,c_{{3}}{c_{{0}}}^{
2}-30\,c_{{1}}c_{{3}}+5\,{c_{{1}}}^{3}-5\,{c_{{0}}}^{4}c_{{1}}
\right) \frac{{z}^{5}}{120} +\ldots
\end{gathered}\end{equation*}

In the case $r_2=1$ the cone of the problem is
$\mathcal{K}=\{k>1\}$. Consequently there are two critical numbers
$k_2=2,\, k_3=3$. Likewise the previous case we find the family of
expansions $G_1^{(0)}2$
\begin{equation}
\label{3.11}w(z)=c_1z + c_2z^2 + c_3z^3 + \sum^{\infty}_{k=4} c_kz^k
\end{equation}
that is generated by the power asymptotic form
$\mathcal{F}_1^{(0)}2: \,\,\, w= c_1\,z$. Here $c_1\neq 0,\, c_2$
and $c_3$ are the arbitrary constants.

For the root $r_2=2$ the cone of the problem is
$\mathcal{K}=\{k>2\}$. It contains $k_3=3$ that is the unique
critical number. The power expansion corresponding to the
asymptotic form $\mathcal{F}_1^{(0)}3: \,\,\, w= c_2\,z^{2}$ takes
the form
\begin{equation}
\label{3.12}w(z)=c_2z^2 +c_3z^3 + \sum^{\infty}_{k=4} c_kz^k
\end{equation}
Again $c_2\neq 0,\,c_3$ are the arbitrary constants. Denote this
family as $G_1^{(0)}3$.

For the root $r_3=3$ the cone of the problem is
$\mathcal{K}=\{k>3\}$. There are no critical numbers in this case.
The expansion of the solution $G_1^{(0)}4$ corresponding to power
asymptotic form $\mathcal{F}_1^{(1)}4: \,\,\, w= c_3\,z^{3}$ can
be written as
\begin{equation}
\label{3.13}w(z)=c_3z^3 + \sum^{\infty}_{k=4} c_kz^k
\end{equation}

Note, that expansions \eqref{3.11}, \eqref{3.12}, \eqref{3.13} are
the special cases of expansion \eqref{3.10}.

Obtained expansions converge for sufficiently small $|z|$. The
existence and the analyticity of expansions \eqref{3.10},
\eqref{3.11}, \eqref{3.12}, \eqref{3.13} follow from the Cauchy
theorem. This apex does not define non-power asymptotic forms and
exponential additions.

\section{Solutions, corresponding to edge $\Gamma_1^{(1)}$.}

Edge $\Gamma_1^{(1)}$ is characterized by the reduced equation
\begin{equation}
\label{4.1}\hat{f}^{(1)}_1 (z,w) \stackrel{def}{=}w_{zzzz}
-5\,w^2\,w_{zz}+5w_zw_{zz} -5\,w\,w_{z}^2+\,w^5=0
\end{equation}
and the normal cone $U^{(1)}_1=\{-\lambda(1,-1),\,\lambda>0\}$.
Therefore $\omega=-1$, i.e. $z\rightarrow 0$ and $r=-1$.
Consequently the solution of equation \eqref{4.1} should be looked
for in the form \begin{equation}\label{e:4-1}w=c_{-1}
z^{-1}\end{equation}
 From the determining equation
\begin{equation}
\label{4.2}c_{-1}(c_{-1}^4 -15\,c_{-1}^2 -10c_{-1}+24) =0
\end{equation}
we find the values of coefficient $c_{-1}$ (not equal to zero):
$c_{-1}^{(1)} =1,\, c_{-1}^{(2)}=-2,\, c_{-1}^{(3)}=-3,\,
c_{-1}^{(4)}=4$. Hence we have four families of power asymptotic
forms
\begin{align}
\label{e4.3_1}&&\mathcal{F}_1^{(1)}1: &\qquad w=z^{-1}
\\&&\label{e4.3_2}\mathcal{F}_1^{(1)}2: &\qquad w=-2z^{-1}
\\
&&\label{e4.4_1}\mathcal{F}_1^{(1)}3: &\qquad
w=-3z^{-1}\\&&\label{e4.4_2}\mathcal{F}_1^{(1)}4: &\qquad w=4z^{-1}
\end{align}
Denote the families of expansions, corresponding to these
asymptotic forms as $G_{1}^{(1)}i,\quad i=1,2,3,4$.

Let us compute the corresponding critical numbers. The first
variation
\begin{equation}
\label{4.5}\frac{\delta f_2^{(1)}}{\delta w}=\frac{d^4}{dz^4}
-10w_{zz}w -5w^2\frac{d^2}{dz^2}
+5w_z\frac{d^2}{dz^2}+5w_zz\frac{d}{dz}-5w_{z}^2 -5\,ww_z \frac
d{dz} + 5\,w^4
\end{equation}
applied to solutions \eqref{e:4-1} yields to operator
\begin{equation}
\label{4.6}\mathcal{L}^(z)
=\frac{d^4}{dz^4}-\frac{5c_{-1}(1+c_{-1})}{z^2}\frac{d^2}{dz^2}+\frac{10c_{-1}(1+c_{-1})}{z^3}
\frac{d}{dz} - \frac{5{c_{-1}}^2(5-{c_{-1}}^2)}{z^4}
\end{equation}

Its characteristic polynomial is
\begin{equation}
\begin{aligned}
\label{4.7}\nu(k) =&{k}^{4}-6\,{k}^{3}+ \left(
11-5\,c_{-1}-5\,{c_{-1}}^{2} \right) {k}^{2}-\\&- \left(
6-15\,c_{-1}-15\,{c_{-1}}^{2} \right) k-5\,{c_{-1}}^{2} \left(
5-{c_{-1}}^{2} \right)
\end{aligned}
\end{equation}
Equation $\nu(k)=0$ has the roots:

1) $k_1=-2,\,k_2=1,\, k_3=2,\,k_4=5$ for $c_{-1}=1 $;

2) $ k_1=-2,\,k_2=1,\, k_3=2,\,k_4=5$ for $c_{-1}=-2 $;

3) $k_1=-2,\,k_2=-3,\, k_3=5,\,k_4=6$ for $c_{-1}=-3 $;

4) $k_1=-2,\,k_2=-8,\, k_3=5,\,k_4=11$ for $c_{-1}=4$ .

 The cone of the
problem here is
\begin{equation}
\label{4.10} \mathcal{K}=\{k>-1\}.
\end{equation}
Thus for  power asymptotic forms \eqref{e4.3_1} and \eqref{e4.3_2}
there are three critical numbers (three roots of the
characteristic polynomial belong to the cone of the problem) and
for the power asymptotic forms \eqref{e4.4_1} and \eqref{e4.4_2}
there are only two critical numbers.

The lattice, generated by basic vectors (-4,1), (1,1) and vector
(-1,-1), corresponding to the shifted carrier of power asymptotic
forms (\ref{e4.3_1} -- \ref{e4.4_2}), consists of points $Q=
m(-4,1)+l(1,1)$. These points intersect with the line $q_{2}=-1$
if $m+l=-1$, i.e. $l=-m-1$. As the cone of the problem here is
\eqref{4.10} then
\begin{equation}
\textbf{K}=\{k=-1+5m,\,\,\,m\in \mathbb{N}\}
\end{equation}
Sets $\textbf{K}(1)$, $\textbf{K}(1,2)$ and $\textbf{K}(1,2,5)$
can be written as
\begin{equation*} \begin{gathered}
\textbf{K}(1)=\{k=-1+5m+2l,\,\,l,m\in
\mathbb{N}\cup\{0\},\,\,l+m\neq\,0\}=\\
=\{1,3,4,5,6,7,8,...\}
\end{gathered}\end{equation*}
\begin{equation*}\begin{gathered}
\textbf{K}(1,2)=\{k=-1+5m+2l+3n,\,l,m,n\in
\mathbb{N}\cup\{0\},\,\,l+m+n\neq\,0\}=\\
=\{1,2,3,4,5,6,7,8,...\}
\end{gathered} \end{equation*}
\begin{equation*}\begin{gathered}
\textbf{K}(1,2,5)=\{k=-1+5m+2l+3n+6j,\,j,l,m,n\in
\mathbb{N}\cup\{0\} ,\,\,j+l+m+n\neq\,0\}=\\
=\{k  ,\,\,k \in \mathbb{N}\}
\end{gathered}\end{equation*}
In this case expansion generated by family \eqref{e4.3_1} takes
the form
\begin{equation}\begin{gathered}
\label{4.15}w(z)=\frac{1}{z}+ \sum^{}_{k\,\in\,\mathbb{N}}
c_{k}^{(1)}\,z^{k}
\end{gathered}\end{equation}

The critical number $1$ does not belong to the set $\textbf{K}$
that is why the compatibility condition for $c_1$ holds
automatically and $c_1$ is the arbitrary constant. The critical
number $2$ also does not belong to the sets $\textbf{K}$,
$\textbf{K}(1)$ and thus $c_2$ is the arbitrary constant. However
the critical number $5$ lies in the sets $\textbf{K}(1)$ and
$\textbf{K}(1,2)$. As a result it is necessary to verify that the
compatibility condition for $c_5$ is true. It holds, so $c_5$ is
the arbitrary constant too.

The three-parametric power expansion that corresponds to power
asymptotic form \eqref{e4.3_1} is as follows
\begin{equation*}\begin{gathered}
w(z)=\frac{1}{z}+c_{{1}}z+c_{{2}}{z}^{2}+\frac14\,{c_{{1}}}^{2}{z}^{3}-\frac{1}{36}
\left( 1+\alpha \right) {z}^{4}+c_{{5}}{z}^{5}+\\+{\frac
{1}{2880}}\,c_{{ 1}} \left( 90\,c_{{2}}c_{{1}}-7-25\,\alpha
\right) {z}^{6}+\\+ \left( \frac16
\,c_{{1}}c_{{5}}+\frac{1}{18}\,{c_{{2}}}^{2}c_{{1}}+{\frac
{1}{192}}\,{c_{{1}} }^{4}+{\frac {1}{540}}\,c_{{2}} \right)
{z}^{7} +\ldots
\end{gathered}
\end{equation*}
where $c_i\equiv c_{i}^{(1)},\quad i=1,2,5$ are the arbitrary
constants.

In the same way we obtain that asymptotic form \eqref{e4.3_2} also
corresponds to three-parameter power expansion
\begin{equation}\begin{gathered}
w(z)=-\frac{2}{z}+ \sum^{}_{k\,\in\,\mathbb{N}} c_{k}^{(2)}\,z^{k}
\end{gathered}\end{equation}
Taking into account seven terms it can be presented as
\begin{equation*}\begin{gathered}
w(z)=-\frac{2}{z}+c_{{1}}z+c_{{2}}{z}^{2}-\frac72\,{c_{{1}}}^{2}{z}^{3}+
\left(
\frac{1}{18}-\frac{1}{36}\,\alpha-\frac{5}{2}\,c_{{1}}c_{{2}}
\right) {z}^{4}+c_{{5}}{ z}^{5}+\\+{\frac {1}{1440}}\,c_{{1}}
\left( 7560\,c_{{1}}c_{{2}}-71+40\, \alpha \right) {z}^{6}+
\\+\left( -\frac23\,c_{{1}}c_{{5}}-{\frac {89}{24}}\,
{c_{{1}}}^{4}-{\frac {19}{540}}\,c_{{2}}+{\frac
{14}{9}}\,{c_{{2}}}^{2 }c_{{1}}+{\frac {1}{54}}\,c_{{2}}\alpha
\right) {z}^{7}
 +\ldots
\end{gathered}
\end{equation*}
where $c_i\equiv c_{i}^{(2)},\quad i=1,2,5$ are the arbitrary
constants.

Sets $\textbf{K}(5)$, $\textbf{K}(5,6)$ and $\textbf{K}(5,11)$ are
as follows
\begin{equation}\begin{gathered}
\textbf{K}(5)=\{k=-1+5m+6l,\,l,m\in
\mathbb{N}\cup\{0\},\,\,l+m\neq\,0\}=\\
=\{4,5,9,10,11,14,15,16,17,19,...\}
\end{gathered}\end{equation}
\begin{equation}\begin{gathered}
\textbf{K}(5,6)=\{k=-1+5m+6l+7n,\,m,l,n\in
\mathbb{N}\cup\{0\},\,\,l+m+n\neq\,0\}=\\
=\{4,5,6,9,10,11,12,13,14,...\}
\end{gathered}\end{equation}
\begin{equation}\begin{gathered}
\textbf{K}(5,11)=\{k=-1+5m+6l+12k,\,l,m,n\in
\mathbb{N}\cup\{0\},\,\,l+m+n\neq\,0\}=\\
=\{4,5,9,10,11,14,15,16,17,19,...\}
\end{gathered}\end{equation}

Using the discussion as described above, we obtain two-parametric
family $G_{1}^{(1)}3$
\begin{equation}\begin{gathered}
w(z)=-\frac{3}{z}+ \sum^{}_{k\,\in\,\textbf{K}(5,6)}
c_{k}^{(3)}\,z^{k}
\end{gathered}\end{equation}
that can be presented as
\begin{equation*}\begin{gathered}
w(z)=-\frac{3}{z}-\left( \frac{1}{28}-{\frac {1}{84}}\,\alpha
\right) {z}^{4}+c_{{5}}{z}^{5}+c_{{6}}{z}^{6}-\\-{\frac
{1}{931392}}\, \left( \alpha-3 \right)  \left( 40 \,\alpha-127
\right) {z}^{9}+{\frac {1}{131040}}\,c_{{5}} \left( 2307
-755\,\alpha \right) {z}^{10}- \\-\left( {\frac
{31}{156}}\,{c_{{5}}}^{2 }-{\frac {547}{38220}}\,c_{{6}}+{\frac
{3}{637}}\,c_{{6}}\alpha
 \right) {z}^{11}
+\ldots
\end{gathered}\end{equation*}
where $c_i\equiv c_i^{(3)},\,i=5,6$ are the arbitrary constants,
and two-parametric family $G_{1}^{(1)}4$
\begin{equation}\begin{gathered}
w(z)=\frac{4}{z}+ \sum^{}_{k\,\in\,\textbf{K}(5,11)}
c_{k}^{(4)}\,z^{k}
\end{gathered}\end{equation}
that can be written as
\begin{equation*}\begin{gathered}
w(z)=\frac{4}{z} +\left( {\frac {1}{126}}+{\frac {1}{504}}\,\alpha
\right) {z}^{4}+c_{{5}}{z}^{5}+\\+{\frac {1}{47500992}}\, \left(
\alpha+4 \right)  \left( 25 \,\alpha+37 \right) {z}^{9}+{\frac
{1}{15120}}\,c_{{5}} \left( 6+5\, \alpha \right)
{z}^{10}+\\+c_{{11}}{z}^{11}+{\frac {1}{2106763997184}}\,
 \left( \alpha+4 \right)  \left( 220\,{\alpha}^{2}+1235\,\alpha+1126
 \right) {z}^{14}
+\ldots
\end{gathered}\end{equation*}
where $c_i\equiv c_i^{(4)},\,i=5,11$ are the arbitrary constants
too.

 Obtained expansions converge
for sufficiently small $|z|$ and have no exponential additions.
The reduced equation \eqref{4.1} also does not have non-power
solutions.

\section{Solutions, corresponding to edge $\Gamma_2^{(1)}$.}

Edge $\Gamma_2^{(1)}$ is characterized by the reduced equation
\begin{equation}
\label{5.1}\hat{f}_2^{(1)} (z,w) \stackrel{def}{=} w^5 - zw=0
\end{equation}
and normal cone
$U^{(1)}_2=\{\lambda(4,1)=4\lambda(1,1/4),\,\lambda>0\}$. It means
that $r=1/4$, $\omega=1$, $z\rightarrow  \, \infty$ and  solution
of this reduced equation is $w=c_{1/4}z^{1/4}$. Substitution this
expression into equation \eqref{5.1} and cancellation the result
by $z^{5/4}$ yields to determining equation $
c_{1/4}(c_{1/4}^4-1)=0$. Thus we have four families of power
asymptotic forms
\begin{align}
\label{5.2} & \mathcal{F}_2^{(1)}1:\,\,  w=
c_{1/4}^{(1)}\,z^{1/4},\quad  c_{1/4}^{(1)}=1
\\
\label{5.3} & \mathcal{F}_2^{(1)}2:\,\, w=c_{1/4}^{(2)}\,
z^{1/4},\quad  c_{1/4}^{(2)}\,=-1
\\
\label{5.4}& \mathcal{F}_2^{(1)}3: \,\,
w=c_{1/4}^{(3)}\,z^{1/4},\quad  c_{1/4}^{(3)}\,=i
\\
\label{5.5} & \mathcal{F}_2^{(1)}4:\,\, w=
c_{1/4}^{(4)}\,z^{1/4},\quad c_{1/4}^{(4)}=-i
\end{align}
Since the reduced equation \eqref{5.1} is algebraic one and the
roots of the determining equation are simple then asymptotic forms
do not have proper (and consequently  critical) numbers and
$\nu(k)\equiv const\neq0$.

The shifted carrier of the power asymptotic forms \eqref{5.2} --
\eqref{5.5} gives a vector $(1/4,-1)$. Points belonging to the
lattice generated by this vector and the basis vectors (-4,1),
(1,1) are the following $Q=(q_1,q_2) =m(1,\,1) +l(1/4,\,-1)
=\left(m+l/4,\,\,m-l \right)$ where $m$, $l$ are whole numbers. At
the line $q_2=-1$ we have $q_1=-1+5/4l$. Taking into consideration
that the cone of the problem is $\mathcal{K}=\{k<1/4\}$ we find
the set $\mathbf{K}$
\begin{equation}
\label{5.6}\mathbf{K}=\{k=-1-5l/4,\,\,\, l\in \mathbb{N}\cup\{0\}
\}
\end{equation}
The expansions to solutions can be written as
\begin{equation}
\label{5.7}G_2^{(1)} n:\,\,\,\,
w(z)=\varphi^{(n)}(z)=c^{(n)}_{1/4} z^{1/4} + \sum^{\infty}_{l=0}
c^{(n)}_{-1-5l/4}\, z^{-1-5l/4}
\end{equation}
In this expression coefficients $c^{(n)}_{-1-5l/4}$ can be
sequentially computed. The calculation of the coefficients
$c_{-1}$ yields to $c_{-1}=\alpha/4$. Taking into account five
terms, we obtain
\begin{equation*}
\begin{gathered}
\label{}w(z)=
\varphi^{(n)}(z)=c_{{1/4}}^{(n)}\,{z^{1/4}}+\frac14\,{\frac
{\alpha}{z}}-{\frac {5}{32}}\,{ \frac {(c_{{1/4}}^{(n)})^{3}
\left( 1+{\alpha}^{2} \right) }{{z}^{9/4}}}+
\\+{\frac {5}{256}}\,{\frac
{(c_{{1/4}}^{(n)})^{2} \left( 8\alpha^3+29\alpha+3
\right)}{{z}^{7/2}}}-{\frac {1}{2048}}\,{\frac {c_{{1/4}}^{(n)}
\left( 365\alpha^4+3210\alpha^2+590\alpha+2013
\right)}{{z}^{19/4}}}+\ldots
\end{gathered}
\end{equation*}

The obtained expansions seem to be divergent ones.

Non-power asymptotic forms do not correspond to edge
$\Gamma^{(1)}_2$ but it generates exponential additions which will
be computed later.

\section{Solutions, corresponding to edge $\Gamma_3^{(1)}$.}

Edge $\Gamma_3^{(1)}$ exists if $\alpha\ne0$. It is characterized
by the reduced equation
\begin{equation}
\label{6.1}\hat{f}_3^{(1)} (z,w) \stackrel{def}{=} - zw - \alpha=0
\end{equation}
and the normal cone $U^{(1)}_3=\{\lambda(1,-1),\,\lambda>0\}$. In
this case $r=-1$, $\omega=1$, $z\longrightarrow \, \infty$ and power
asymptotic can be presented in the form
\begin{equation}
\label{6.2}\mathcal{F}_3^{(1)}1:\qquad w= \frac
{c_{-1}}{z},\qquad\,c_{-1}=-\alpha
\end{equation}
As equation \eqref{6.1} is algebraic one and
\begin{equation}
\label{6.3}\nu(k)=z^{-1}\frac{\delta \hat{f}_3^{(1)}}{\delta y}=-1
\end{equation}
then the  solutions  of equation \eqref{6.1} do not have critical
numbers. The cone of the problem is $\mathcal{K}=\{k<-1\}$. The
shifted carrier of the power asymptotic \eqref{6.2} gives the
vector $(-1,-1)$, which belongs to the lattice generated by the
carrier of the studied equation. So we obtain the set $\mathbf{K}$
\begin{equation}
\label{6.4}\mathbf{K}=\{k=-1-5m,\,\,\, m\in \mathbb{N} \}
\end{equation}
Therefore we can determine the power expansion, corresponding to
the asymptotic form \eqref{6.2}
\begin{equation}
\label{6.5}G_3^{(1)}:\,\,\,\, w(z)=\widetilde{\varphi}(z)=\frac
{c_{-1}}{z} + \sum^{\infty}_{m=1} c_{-1-5m}\, z^{-1-5m}
\end{equation}
In this expression all coefficients can be sequentially found.
Taking into account three terms, it can be rewritten as
\begin{equation*}
\begin{gathered}
\label{}w(z)=\widetilde{\varphi}(z)=-{\frac {\alpha}{z}}-{\frac
{P}{{z}^{6}}}-{\frac {P \left( 5\,
{\alpha}^{4}-295\,{\alpha}^{2}+270\alpha +3024 \right)
}{{z}^{11}}}+\ldots
\end{gathered}
\end{equation*}
where
\begin{equation}\label{e:P}P=\alpha(\alpha+1)(\alpha-2)(\alpha-3)(\alpha+4)\end{equation} and
all other not written out coefficients are proportional to this
factor. Hence if $P=0$ this expansion determines the simplest
exact rational solutions of equation \eqref{e:eq}.

Edge $\Gamma^{(1)}_3$ does not generate non-power asymptotic forms
but it defines exponential additions, which will be found below.

\section{Solutions, corresponding to edge $\Gamma_4^{(1)}$.}

Edge $\Gamma_4^{(1)}$ exists, if $\alpha\ne0$. It defines the
following reduced equation
\begin{equation}
\label{7.1}\hat{f}_4^{(1)} (z,w)\stackrel{def}{=}w_{zzzz} - \alpha=0
\end{equation}
and the normal cone $U^{(1)}_4 =\{-\lambda(1,4),\,\,\lambda>0\}$.
Thus $\omega=-1$, i.e. $z\rightarrow 0$, $r=4$ and we have the
unique family of power asymptotic forms
\begin{equation}
\label{7.2}\mathcal{F}_4^{(1)}1: \quad w=c_4z^4,\qquad
c_4=\frac{\alpha}{24}
\end{equation}
Let us find the critical numbers. The first variation of
equation \eqref{7.1} is
\begin{equation}
\label{7.3}\frac{\delta \hat{f}^{(1)}_4}{\delta w} =\frac{d^4}{dz^4}
\end{equation}
The proper numbers are $k_1=0$, $k_2=1$, $k_3=2$, $k_4=3$. None of
them belongs to the cone of the problem $ \mathcal{K}=\{k>4\}$.
Consequently  there are no critical numbers here.

Vector, corresponding to the shifted carrier of asymptotic form
\eqref{7.2} is (4,-1), so it belongs to the lattice generated by
the basis vectors. Therefore set $\textbf{K}$ is
\begin{equation}
\label{7.4}\textbf{K}=\{k=4+5m,\quad m\in \mathbb{N}\}
\end{equation}
and the power expansion can be written as
\begin{equation}
\label{7.5}w(z)=z^4\left(\frac{\alpha}{24}+\sum^{\infty}_{m=1}
c_{4+5m}\,z^{5m}\right)
\end{equation}
All the coefficients can be uniquely computed. Taking into account
three terms, we can write this expansion $G_4^{(1)}1$ as
\begin{equation*}
w(z)=\frac{1}{24}\alpha\,{z}^{4}-{\frac {\left( 10\,\alpha-1
 \right)}{72576}}\,\alpha\,
 {z}^{9}+{\frac {\left( 3120\,{\alpha}^{2}-185\,\alpha
+2 \right)}{3487131648}}\,\alpha\,  {z}^{14}+\ldots
\end{equation*}

The obtained expansion can be considered as the special case of
expansion \eqref{3.10} at $c_0=c_1=c_2=c_3=0$. It converges for
sufficiently small $|z|$.

 Edge $\Gamma^{(1)}_4$ does not define
exponential additions and non-power asymptotic forms.

\section{Solutions, corresponding to edge $\Gamma^{(1)}_5$.}

Edge $\Gamma^{(1)}_5$ exists, if and only if $\alpha=0$.

The reduced equation which corresponds to the edge $\Gamma^{(1)}_5$, takes the form
\begin{equation}
\label{11.1}\hat{f}_2^{(5)} (z,w) \stackrel{def}{=} w_{zzzz}-zw=0
\end{equation}
It does not possess solutions in the form $w=c_{r}z^{r}$, except
of the trivial one $w\equiv 0$. But edge $\Gamma^{(1)}_5$ defines
non-power asymptotic forms of equation \eqref{e:eq}.

As edge $\Gamma^{(1)}_5$ is the horizontal one, we can use the
logarithmic transformation
\begin{equation}
\label{11.2}y=\frac{d \ln w}{dz}
\end{equation}
Hence we have the relations
\begin{equation*}
\begin{gathered}
\label{}w'=yw,\,\,\,w''=(y'+y^2)w,\,\,\,w'''=(y''+3yy'+y^3)w,\\
w''''=(y'''+4yy''+3{y'}^2+6y^2y'+y^4)w
\end{gathered}
\end{equation*}
After application this transformation and after cancellation of
the result by $w$   equation \eqref{11.1} can be rewritten as
\begin{equation}
\label{11.3}h(z,y)\stackrel{def}{=}y'''+4yy''+3{y'}^2+6y^2y'+y^4-z=0
\end{equation}

The carrier $S(h)$ of this equation consists of the following
points: $M_1=(-3,1)$, $M_2=(0,4)$, $M_3=(1,0)$, $M_5=(-2,2)$ and
$M_6=(-1,3)$.

\begin{figure}[h] 
\center%
\epsfig{file=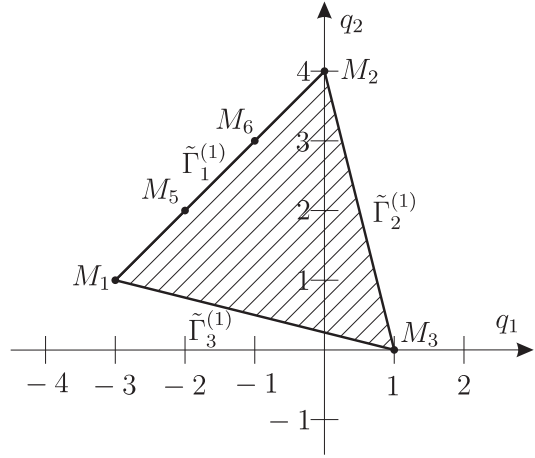,width=70mm} \caption{The carrier of equation
\eqref{11.3}.}\label{fig:ff5}
\end{figure}

\begin{figure}[h] 
\center%
\epsfig{file=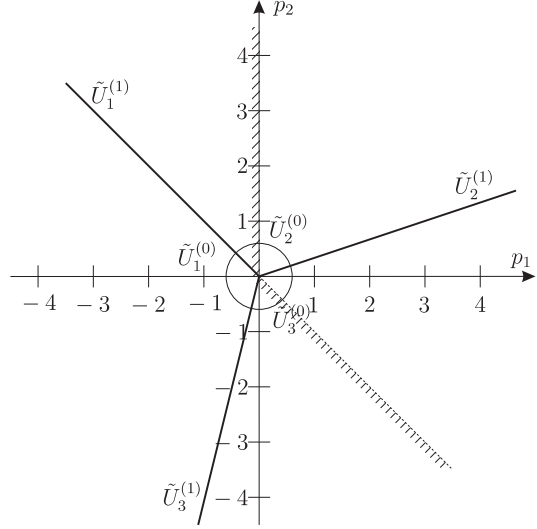,width=70mm} \caption{The normal cones of
equation \eqref{11.3}.}\label{fig:ff6}
\end{figure}

Their convex hull $\Gamma(h)$ is the triangle (fig. \ref{fig:ff5}).
The normal cones of its bounds are presented at fig. \ref{fig:ff6}.

According to \cite{Bruno03}, the cone of the problem here is
$p_1+p_2\ge0$, $p_1\ge0$. It intersects with the normal cones
$\tilde{U}_2^{(0)}$, $\tilde{U}_3^{(0)}$, $\tilde{U}_2^{(1)}$
only. So we should examine the reduced equations which correspond
to the bounds $\tilde{\Gamma}_2^{(0)}$, $\tilde{\Gamma}_3^{(0)}$,
$\tilde{\Gamma}_2^{(1)}$.

Apexes $\tilde{\Gamma}_2^{(0)}$ and $\tilde{\Gamma}_3^{(0)}$ are
characterized by the algebraic reduced equations $y^4=0$ and
$-z=0$ accordingly, so they do not give suitable solutions.

 Edge $\tilde{\Gamma}_2^{(1)}$ defines algebraic
reduced equation
\begin{equation}
\label{11.4}y^4-z=0
\end{equation}
which has four suitable solutions $y^{(n)}(z)=c_{1/4}^{(n)}
z^{1/4}$, where $c_{1/4}^{(1,2)}=\pm1$, $c_{1/4}^{(3,4)}=\pm i$,
$r=1/4$ and $\omega=1 \Rightarrow z\rightarrow\infty$. These
solutions do not have critical numbers.

Using the inversion to transformation \eqref{11.2}, we obtain four
weak asymptotic forms for solutions of equation \eqref{e:eq} near
$z\rightarrow\infty$
$$w(z)=C_1\exp \left(\frac45 c_{1/4}^{(n)} z^{5/4}\right)$$
where $n=1,2,3,4$ and $C_1$ is the arbitrary constant.

Now let us obtain the strong asymptotic forms.

The lattice of the shifted carrier of equation \eqref{11.3} is
generated by vectors $(1/4,-1)$ and $(1,1)$. Taking into account
the cone of the problem $\mathcal{K}=\{k<1/4\}$, we find set
$\textbf{K}$
\begin{equation}
\label{11.5}\textbf{K}=\{k=1/4-5l/4,\quad l\in \mathbb{N}\}
\end{equation}
So equation \eqref{11.3} has solutions in the form
\begin{equation}
\label{11.6}y^{(n)}(z)=c^{(n)}_{1/4} z^{1/4} + \sum^{\infty}_{l=1}
c^{(n)}_{1/4-5l/4}\, z^{1/4-5l/4},\,\,\,n=1,2,3,4
\end{equation}
where the coefficients $c^{(n)}_{1/4-5l/4}$ are uniquely defined.
The coefficient $c^{(n)}_{-1}$ does not depend on $n$:
$c^{(n)}_{-1}=-3/8$. Taking into account transformation
\eqref{11.2}, we get
\begin{equation}
\label{11.7}\ln w=\int ydz=
\frac45c^{(n)}_{1/4}z^{5/4}+C_0-\frac38\ln z+\sum^{\infty}
_{l=2}\frac{4c^{(n)}_{1/4-5l/4}}{5(1-l)}z^{5(1-l)/4}
\end{equation}
In this expression $C_0$ is a constant of integration. So we
obtain
\begin{equation}
\begin{gathered}
\label{11.8}w(z)=\frac{C_1}{z^{3/8}}\exp\left[\frac45c^{(n)}_{1/4}z^{5/4}+\sum^{\infty}
_{l=2}\frac{4c^{(n)}_{1/4-5l/4}}{5(1-l)}z^{5(1-l)/4}\right]
\end{gathered}
\end{equation}
Taking into account three terms of the series, the obtained
non-power strong asymptotic forms of equation \eqref{e:eq},
existing under $\alpha=0$, can be written as
\begin{equation*}
\begin{gathered}
\label{}w(z)=\frac{C_1}{z^{3/8}}\exp\left[\frac45c^{(n)}_{1/4}z^{5/4}+{\frac
{9}{32}}\,{\frac {1}{c^{(n)}_{1/4}{z}^{5/4} }}+{\frac
{45}{256}}\,\frac {1}{(c^{(n)}_{1/4})^{2}{z}^{5/2}}+\ldots \right]
\end{gathered}
\end{equation*}

\section{Exponential additions of the first level for expansions, corresponding to edge $\Gamma^{(1)}_2$.}

Let us find exponential additions of the first level for
expansions \eqref{5.7}, corresponding to edge $\Gamma^{(1)}_2$.

We are looking for solutions in the form
\begin{equation}
\label{13.1}w(z)=\varphi(z)+u(z)
\end{equation}
Here and later we use the notation $\varphi=\varphi^{(n)}$,
$u=u^{(n)}$, $n=1,2,3,4$.

The reduced equation for the addition $u(z)$ is a linear equation
\begin{equation}
\label{13.2}M^{(1)}(z) u(z)=0
\end{equation}
where $M^{(1)}(z)$ is the first variation of equation \eqref{e:eq}
at solution $w(z)=\varphi(z)$. As long as
\begin{equation}
\label{13.3}\frac{\delta f}{\delta w} =\frac{d^4}{dz^4}
-5w^{2}\frac{d^2}{dz^2} - 10 w w_{zz}+5w_z\frac{d^2}{d
z^2}+5w_{zz}\frac{d}{d z}-10 ww_z \frac{d}{dz} -5w_z^{2}+5w^4-z
\end{equation}
then
\begin{equation}
\label{13.4}M^{(1)}(z) =\frac{d^4}{dz^4}
-5(\varphi^{2}-\varphi_z)\frac{d^2}{dz^2}+5(\varphi_{zz}-2\varphi
\varphi_z)\frac{d}{d z} -5\varphi_z^{2} - 10 \varphi
\varphi_{zz}+5\varphi^4-z
\end{equation}
and equation \eqref{13.2} can be rewritten as
\begin{equation}\label{13.5}
\begin{aligned}
\frac{d^4 u}{dz^4} -5(\varphi^{2}-\varphi_z)\frac{d^2
u}{dz^2}&+5(\varphi_{zz}-2\varphi \varphi_z)\frac{d u}{d z}-\\
-&[5\varphi_z^{2} + 10 \varphi \varphi_{zz}-5\varphi^4+z]u=0
\end{aligned}
\end{equation}

Assumed that
\begin{equation}
\label{13.6}\zeta(z)=\frac{d \ln u(z)}{dz}
\end{equation}
we have
\begin{equation*}\begin{gathered} \frac{d u}{dz} =\zeta u,
\qquad \frac{d^2u}{dz^2 }=\zeta _z u + \zeta^2 u, \qquad
\frac{d^3u}{dz^3} =\zeta_{zz} u+ 3 \zeta  \zeta_z u + \zeta^3 u,
\\
\frac{d^4u}{dz^4} =\zeta_{zzz} u+ 4 \zeta \zeta_{zz} + 3 \zeta_z^2
u + 6 \zeta^2 \zeta _z u + \zeta^4 u
\end{gathered}
\end{equation*}

Using these expansions, from equation \eqref{13.5} we obtain
\begin{multline}\label{13.7}
\zeta_{zzz}+4\zeta\zeta_{zz}+3\zeta_z^2+6\zeta^2\zeta_z+\zeta^4
-5(\varphi^2-\varphi_z)(\zeta_z+\zeta^2)+\\+5\zeta(\varphi_{zz}-2\varphi\varphi_z)
+5\varphi^4-5\varphi_z^2-10\varphi\varphi_{zz}-z=0
\end{multline}

The carrier of this equation consists of points
\begin{equation}
\begin{gathered}
\label{13.8}Q_{k,0} =\left(1-\frac{5}{4}k,0\right),\qquad
Q_{k,1}=\left(-\frac{1}{2}-\frac{5}{4}k,1\right),\\ Q_{k,2}=
\left(\frac{1}{2}-\frac{5}{4}k,2\right), \qquad Q_{0,3}=(-1,3),\qquad Q_{0,4}=(0,4),\\
 k \in \mathbb{N}\cup\{0\}
\end{gathered}
\end{equation}
where the doubled index in the points numeration is introduced for the notation
convenience.

\begin{figure}[h] 
\center%
\epsfig{file=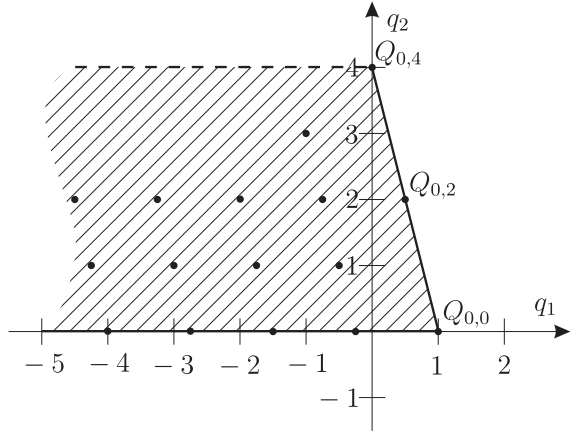,width=75mm} \caption{The carrier of equation
\eqref{13.7}}\label{fig:ff7}
\end{figure}

The convex hull of these points is the string, presented at fig.
\ref{fig:ff7}.

It should examine the edge, passing through points
$Q_{0,0}=(1,0)$, $Q_{0,2}=(1/2,2)$ and $Q_{0,4}=(0,4)$, the
external normal here is $N=(4,1)$. This edge is corresponded by
reduced equation
\begin{equation}
\label{13.9}h_1^{(1)}(z,\zeta) \stackrel{def}{=} \zeta^4
-5c_{1/4}^2 \sqrt{z}\zeta^2 + (5c_{1/4}^4-1)z=0
\end{equation}
where $c_{1/4}$ is the highest coefficient of expansion
$\varphi(z)$ and can have four possible values $c_{1/4}^{(n)}$,
$n=1,2,3,4$ (see \eqref{5.2}-\eqref{5.5}). Taking into account,
that $\forall n \Rightarrow (c_{1/4}^{(n)})^4=1 $, for each value
of $c_{1/4}$ we obtain four solutions of equation \eqref{13.9}
\begin{equation}
\label{13.10}\zeta(z) =g_{1/4}z^{1/4}
\end{equation}
where
\begin{equation*}
g_{1/4}\equiv g_{1/4}^{(n;\,\, l=\{1,2,3,4\})}=\pm
c_{1/4}^{(l)}\sqrt{\frac{5\pm3}2}
\end{equation*}
Reduced equation \eqref{13.9} is algebraic one, so it does not
have critical numbers.

The shifted carrier of reduced solutions \eqref{13.10} gives the
vector $\left(-1/4,1 \right)$, which belongs to the lattice,
generated by the points of the  carrier of equation \eqref{13.7}.
The basis vectors of this lattice are $(5/4,0)$ and $(1,1)$. So we
can present the points of this lattice in the form
\begin{equation*}
Q=(q_1,q_2)=k(1,1)
+m\left(\frac54,0\right)=\left(k+\frac{5m}4,k\right)
\end{equation*}
At the line $q_2=-1$ we have $k=-1$ and hence $q_1=-1+5m/4$. Since
the cone of the problem is $\mathcal{K}=\left\{k<\frac14\right\}$,
set $\mathbf{K}$ can be written as
\begin{equation}
\label{13.11}\mathbf{K}=\left\{\frac{1-5k}4,k\in\mathbb{N}\right\}
\end{equation}
Power expansions to solutions of equation \eqref{13.7} are
\begin{equation}
\label{13.12}\zeta(z)\equiv \zeta^{(n,\,l)}(z)=g^{(n,\,l)}_{1/4}
z^{1/4}
+\sum_{k\,\in\,\,\mathbb{N}}g^{(n,\,l)}_{(1-5k)/4}\,z^{(1-5k)/4},\qquad
 n,l=1,2,3,4
\end{equation}
where coefficients $g^{(n,\,l)}_{(1-5k)/4}$, $k \in \mathbb{N}$
can be uniquely determined.

For example, coefficient $g_{-1}^{1/4}$ can be expressed as
$$g_{-1}^{1/4}={\frac {10 \alpha s
({s}^{2}-2)-5\,{s}^{3}-15\,{s}^{ 2}+24}{8(5{s}^{2}-8)}},\qquad
s=\frac{g_{1/4}}{c_{1/4}}
$$
that for different values of $g_{1/4}^{(n)}$ gives values
$(5\alpha-2)/12$, $(-5\alpha-7)/12$, $(10\alpha-19)/24$ or
$(-10\alpha+1)/24$.

Using the inverse transformation to \eqref{13.6}
\begin{equation}
\label{13.13}u(z) = C_1 \exp \int \zeta(z) dz
\end{equation}
we can compute the exponential additions
\begin{equation}
\begin{gathered}
\label{13.14}u(z)\equiv u^{(n,\,l)}(z)=C_1\,z^{g^{(n,\,l)}_{-1}}\,
\exp \left[\frac45\, g^{(n,\,l)}_{1/4}\,z^{5/4} +
\sum^{\infty}_{k=2} \frac{4}{5(1-k)} g^{(n,\,l)}_{(1-5k)/4}
z^{5(1-k)/4}\right]\\
 n,l=1,2,3,4
\end{gathered}
\end{equation}
Here $C_1$ and later $C_2$ and $C_3$ are constants of integration.

So for each expansion $G^{(1)}_{2}n$ we have found four
one-parametric families of additions $G_2^{(1)}n\, G^{(1)}_1 l$
($n,l=1,2,3,4$).

Additions $u^{(n,\,l)}(z)$ are exponentially small at
$z\rightarrow \infty$ in those sectors of the complex plane $z$,
where
\begin{equation}
\label{13.15}Re \left[g^{(n,\,l)}_{1/4}\, z^{5/4}\right]<0
\end{equation}

\section{Exponential additions of the second level, corresponding to edge $\Gamma^{(1)}_2$.}

In this section we compute the exponential additions of the second
level $v(z)$, i.e. the additions to the solutions $\zeta(z)$. The
reduced equation for this addition takes the form
\begin{equation}
\label{14.1}M^{(2)} (z) v(z)=0
\end{equation}
where operator $M^{(2)}(z)$ is the first variation of \eqref{13.7}
at  solutions $\zeta(z)$. Equation \eqref{14.1} can be rewritten
as
\begin{multline}
\label{14.2}\frac{d^3v}{dz^3} + 4\zeta \frac{d^2v}{dz^2} +
(6\zeta_z + 6\zeta^{2}-5\varphi^2+5\varphi_z)\frac{dv}{dz} +\\+
(4\zeta_{zz} +12\zeta\zeta_z + 4\zeta^3 - 10\varphi^2\zeta
+10\varphi_z\zeta - 10\varphi\varphi_z+5\varphi_{zz})v=0
\end{multline}

 Making the transformation of variables
\begin{equation}
\label{14.3}\frac{d \ln v}{dz}=\xi,
\end{equation}
we have
\begin{equation*}
\label{}\frac{dv}{dz}=\xi v,\,\,\quad\, \frac{d^2v}{dz^2} =\xi_z
v+\xi^2 v,\,\,\quad\, \frac{d^3 v}{dz^3}=\xi_{zz} v + 3\xi\xi_{z}
v+\xi^3 v
\end{equation*}
and equation \eqref{14.2} transforms to
\begin{multline}
\label{14.4}\xi_{zz} + 3\xi\xi_{z}+\xi^3 + 4\zeta (\xi_z+\xi^2) +
(6\zeta_z + 6\zeta^{2}-5\varphi^2+5\varphi_z)\xi +\\+ 4\zeta_{zz}
+12\zeta\zeta_z + 4\zeta^3 - 10\varphi^2\zeta +10\varphi_z\zeta -
10\varphi\varphi_z+5\varphi_{zz}=0
\end{multline}

The carrier of this equation consists of points

\begin{equation}
\begin{gathered}
\label{14.5}Q_{0,\,3}=(0,3),\quad
Q_{k,\,2}=\left(\frac14-\frac54k,\,2\right),\quad
Q_{k,\,1}=\left(\frac12-\frac54k,\,1\right),\\
Q_{k,\,0}=\left(\frac34-\frac54k,\,1\right), \qquad k \in
\mathbb{N}\cup\{0\}
\end{gathered}
\end{equation}

\begin{figure}[h] 
\center%
\epsfig{file=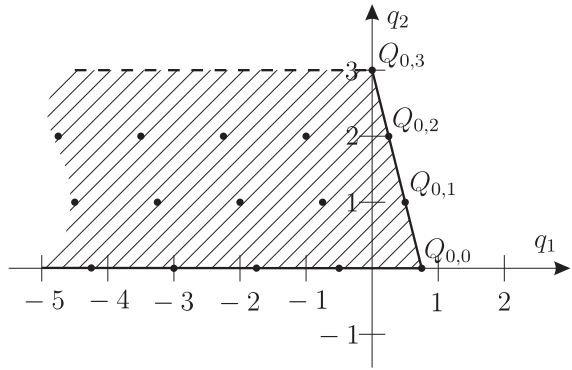,width=75mm} \caption{The carrier of equation
\eqref{14.4}}\label{fig:ff8}
\end{figure}

Their convex hull is the string, presented at fig. \ref{fig:ff8}. To
obtain the exponential additions it should examine the edge, passing
through points $Q_{0,\,3}=(0,3)$, $Q_{0,\,2}=(1/4,2)$,
$Q_{0,\,1}=(1/2,1)$ and $Q_{0,\,0}=(3/4,\,0)$.
 The reduced
equation corresponding to this edge is
\begin{equation}
\begin{gathered}
\label{14.6}\xi^3 + 4g_{1/4}z^{1/4}\xi^2 +
[6g_{1/4}^2-5c_{1/4}^2]z^{1/2}\xi +
[4g_{1/4}^3-10g_{1/4}c_{1/4}^2]z^{3/4} =0
\end{gathered}
\end{equation}

Solution of equation \eqref{14.6} can be presented in the form
\begin{equation}
\label{14.7}\xi(z)=r_{1/4} z^{1/4}
\end{equation}
where $r_{1/4}$ is one of the roots of the cubic equation
\begin{equation}
\begin{gathered}
\label{14.8} r^3 +4\,g_{1/4}r^2+ \left(6\,g_{1/4}^2
-5\,c_{1/4}^2\right)\,r+4\, g_{1/4}^{3} -10\,g_{1/4}\,c_{1/4}^2=0.
\end{gathered}
\end{equation}
and, therefore, also depends on $c_{1/4}=c_{1/4}^{n}$ and
$g_{1/4}^{n,\,l}$, $n,l=1,2,3,4$. Solving this equation we obtain
\begin{equation*}
\label{}r_{1/4}^{(n,\,l,\,m=1)}=-2\,g_{1/4},\quad\,
r_{1/4}^{(n,\,l,\,m=\{2,3\})}=-g_{1/4}\pm\sqrt{{5\,c_{1/4}^2-g_{1/4}^2}}
\end{equation*}

Lattice, corresponding to the shifted carrier of equation
\eqref{14.4}, can be generated by vectors $(1,\,1)$ and
$(5/4,\,0)$. Vector $(-1/4,1)$, conforming to the shifted carrier
of reduced solutions \eqref{14.7}, belongs to this lattice. So set
$\mathbf{K}$ coincides with \eqref{13.11} and power expansions of
functions  $\xi^{(n,\,l,\,m)}(z)$ can be written as

\begin{equation}
\begin{gathered}
\label{14.9}
\xi(z)\equiv\xi^{(n,\,l,\,m)}(z)=r_{1/4}^{(n,\,l,\,m)} z^{1/4}
+\sum_{k\,\in\,\,\mathbb{N}}r^{(n,\,l,\,m)}_{(1-5k)/4}\,z^{(1-5k)/4},\\
\,\quad\,n,l=1,2,3,4;\,\,\quad\, m=1,2,3
\end{gathered}
\end{equation}
where coefficients $r^{(n,\,l,\,m)}_{(1-5k)/4}$, $k\in \mathbb{N}$
can be uniquely determined.

So we have found exponential additions $v^{(n,\,l,\,m)}(z)$ to
 solutions $\zeta^{(n,\,l)}(z)$
\begin{equation}
\begin{gathered}
\label{14.10}v(z)\equiv
v^{(n,\,l,\,m)}(z)=C_2\,z^{r_{-1}^{(n,\,l,\,m)}}\, \exp
\left[\frac45\, r_{1/4}^{(n,\,l,\,m)}\,z^{5/4} +
\sum^{\infty}_{k=2} \frac{4}{5(1-k)} r^{(n,\,l,\,m)}_{(1-5k)/4}
z^{5(1-k)/4}\right]\\
\,\, n=1,2,3,4;\,\quad\,l=1,2,3,4;\,\quad\,m=1,2,3
\end{gathered}
\end{equation}
They are exponentially small at $z\rightarrow\infty$ on condition
that
\begin{equation}
\label{14.11}Re \left[r_{1/4}^{(n,\,l,\,m)}\,z^{5/4}\right]<0
\end{equation}

\section{Exponential additions of the third level, corresponding to edge $\Gamma^{(1)}_2$.}

\begin{figure}[h] 
\center%
\epsfig{file=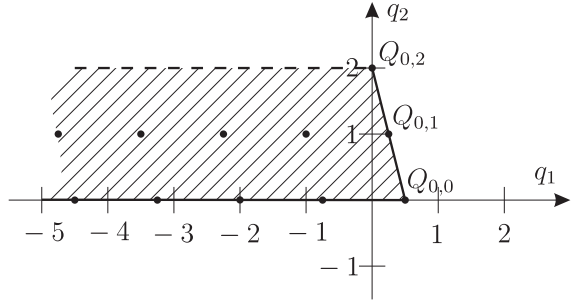,width=75mm} \caption{The carrier of equation
\eqref{15.4}}\label{fig:ff9}
\end{figure}

In this section we are looking for exponential additions of the
third level $\theta(z)$, i.e. additions to the solutions $\xi(z)$.
The reduced equation for addition $\theta(z)$ is the following
\begin{equation}
\label{15.1}M^{(3)} (z) \theta(z)=0
\end{equation}
Operator $M^{(3)}(z)$ can be found as the first variation of
\eqref{14.4} at solutions $\xi(z)$ and then equation \eqref{15.1}
for function $\theta(z)$ can be rewritten as
\begin{equation}
\begin{gathered}
\label{15.2}\theta_{zz} + (3\xi + 4\zeta)\theta_z +
(3\xi_z+3\xi^2+8\xi\zeta+6\zeta_z+6\zeta^2-5\varphi^2+5\varphi_z)\theta=0
\end{gathered}
\end{equation}
Using new variable $\eta$, satisfying the relation
\begin{equation}
\label{15.3}\frac{d \ln \theta}{dz}=\eta
\end{equation}
and taking into account that
\begin{equation*}
\label{}\frac{d\theta}{dz}=\eta \theta,\,\,\quad\,
\frac{d^2\theta}{dz^2} =\eta_z \theta+\eta^2 \theta
\end{equation*}
we obtain, that equation \eqref{15.2} transfers to equation
\begin{equation}\begin{gathered}
\label{15.4}\eta_{z} +\eta^2+ (3\xi +
4\zeta)\eta+3\xi_z+3\xi^2+8\xi\zeta+6\zeta_z+6\zeta^2-5\varphi^2+5\varphi_z=0
\end{gathered}\end{equation}
The carrier of this equation is composed of points
\begin{equation}
\label{15.5}Q_{0,\,2}=(0,2),\quad
Q_{k,\,1}=\left(\frac14-\frac54k,1\right),\quad
Q_{k,\,0}=\left(\frac12-\frac54k,\,1\right),\qquad k \in
\mathbb{N}\cup\{0\}
\end{equation}
The convex hull of these points is the string, presented at fig.
\ref{fig:ff9}. To obtain the exponential additions it should examine
the edge, passing through points $Q_{0,2}=(0,2)$, $Q_{0,1}=(1/4,1)$
and $Q_{0,0}=(1/2,0)$. Reduced equation, corresponding to this edge,
can be written as
\begin{equation}
\begin{gathered}
\label{15.6}\eta^2+ (3r_{1/4}+4g_{1/4})z^{1/4}\eta+
\left(3r_{1/4}^2+8r_{1/4}g_{1/4}+ 6g_{1/4}^2-5c_{1/4}^2
\right)z^{1/2}=0
\end{gathered}
\end{equation}
where $c_{1/4}\equiv c_{1/4}^{(n)}$, $g_{1/4}\equiv
g_{1/4}^{(n,l)}$, $r_{1/4}\equiv r_{1/4}^{(n,l,m)}$,
$n,l=1,2,3,4$, $m=1,2,3$.

Solution of equation \eqref{15.6} can be presented as
\begin{equation}
\label{15.7}\eta(z)=q_{1/4} z^{1/4}
\end{equation}
where $q_{1/4}$ is one of the roots of the quadratic equation
\begin{equation}
\label{15.8}q^2+ (3r_{1/4}+4g_{1/4})q+ 3r_{1/4}^2+8r_{1/4}g_{1/4}+
6g_{1/4}^2-5c_{1/4}^2 =0
\end{equation}
Solving this equation we obtain
\begin{equation*}
\begin{gathered}
\label{}
q_{1/4}^{(n,\,l,\,m,\,p=\{1,\,2\})}=-\frac32\,r_{1/4}-2\,g_{1/4}\pm\frac12\sqrt{{20\,c_{1/4}^2-3\,r_{1/4}^{2}-
8\,r_{1/4}\,g_{1/4}-8\, g_{1/4}^{2}}}
\end{gathered}
\end{equation*}
The lattice of the shifted carrier of equation \eqref{15.4} can be
generated by vectors $(1,\,1)$, and $(5/4,\,0)$. So  set
$\mathbf{K}$ coincides with \eqref{13.11}. Power expansions for
$\eta^{(n,\,l,\,m,\,p)}(z)$ can be presented as
\begin{equation}
\begin{gathered}
\label{15.9}\eta(z)\equiv
\eta^{(n,\,l,\,m,\,p)}(z)=q^{(n,\,l,\,m,\,p)}_{1/4} z^{1/4}
+\sum_{k\,\in\,\,\mathbb{N}}q^{(n,\,l,\,m,\,p)}_{(1-5k)/4}\,z^{(1-5k)/4},\\
n,l=1,2,3,4;\,\, m=1,2,3;\,\, p=1,2
\end{gathered}
\end{equation}
where coefficients $q^{(n,\,l,\,m,\,p)}_{(1-5k)/4}$, $k\in
\mathbb{N}$ can be uniquely determined.

So, exponential additions $y^{(n,\,l,\,m,\,p)}(z)$ to solutions
$\xi^{(n,\,l,\,m)}(z)$ can be written as
\begin{multline}
\label{15.10}\theta(z)\equiv
\theta^{(n,\,l,\,m,\,p)}(z)=\\=C_3\,z^{q_{-1}^{(n,\,l,\,m,\,p)}}
\exp \left[\frac45\, q^{(n,\,l,\,m,\,p)}_{1/4}\,z^{5/4} +
\sum^{\infty}_{k=2} \frac{4}{5(1-k)}
q^{(n,\,l,\,m,\,p)}_{(1-5k)/4}
z^{5(1-k)/4}\right]\\
n,l=1,2,3,4;\,\quad\, m=1,2,3;\,\quad\, p=1,2
\end{multline}
They are exponentially small at $z\rightarrow\infty$ if
\begin{equation}
\label{15.11}Re \left[q_{1/4}^{(n,\,l,\,m,\,p)}\,z^{5/4}\right]<0
\end{equation}

Thus for expansion \eqref{5.7} to solutions of equation
\eqref{e:eq} near $z=\infty$ three levels of exponential additions
have been found. Therefore, solutions $w(z)$ at
$z\rightarrow\infty$ can be written as
\begin{multline*}
\label{}w(z)=\varphi^{(n)}(z)+\\+\exp\left[ \int dz \left(
\zeta^{(n,\,l)}(z)+ \exp \left[\int
dz\left(\xi^{(n,\,l,\,m)}(z)+\exp\left(\int
dz\,\eta^{(n,\,l,\,m,\,p)}(z)\right)\right)\right]\right)\right]
\end{multline*}
where $n,l=1,2,3,4;\,\,\,m=1,2,3;\,\,\,p=1,2$.

\section{Three levels of exponential additions, corresponding to edge $\Gamma^{(1)}_3$.}

Let us find the exponential addition of the first level $u(z)$ to
expansion \eqref{6.5}, i.e.
\begin{equation}
\label{16.1} w(z)=\widetilde{\varphi}(z) + \widetilde{u}(z)
\end{equation}
To obtain this addition we use the technique, described in section
9.

Using new variable
\begin{equation}
\widetilde{\zeta}(z)=\frac{d \ln \widetilde{u}(z)}{dz}
\end{equation}
we obtain equation for $\widetilde{\zeta}(z)$

\begin{equation}
\begin{gathered}
\label{16.2} \widetilde{\zeta}_{zzz} +
4\widetilde{\zeta}{\widetilde{\zeta}}_{zz}
+3{\widetilde{\zeta}}_z^2 +6 {\widetilde{\zeta}}^2
{\widetilde{\zeta}}_z +  {\widetilde{\zeta}}^4
-5({\widetilde{\varphi}}^2-
{\widetilde{\varphi}}_z)( {\widetilde{\zeta}}_z + {\widetilde{\zeta}}^2)+\\
+5{\widetilde{\zeta}}({\widetilde{\varphi}}_{zz}-2{\widetilde{\varphi}}{\widetilde{\varphi}}_z)+5
{\widetilde{\varphi}}^4-5{\widetilde{\varphi}}_z^2-10{\widetilde{\varphi}}{\widetilde{\varphi}}_{zz}
-z=0
\end{gathered}
\end{equation}
The carrier of this equation consists of points
\begin{equation}
\begin{gathered}
\label{16.3}Q_{k,\,0} =\left(1-5k,0\right),\quad Q_{k,\,1}=\left(-3-5k,1\right),\\
Q_{k,\,2}=\left(-2-5k,2\right),\quad Q_{0,\,3}=(-1,3), \quad Q_{0,\,4}=(0,\,4),\\
k \in \mathbb{N}\cup\{0\}
\end{gathered}
\end{equation}

The convex hull of these points is the string, similar to one,
presented at fig. \ref{fig:ff7}.

The reduced equation, corresponding to the edge, passing through
points $Q_{0,\,0}=(1,0)$ and $Q_{0,\,4}=(0,4)$, can be written as
\begin{equation}
\label{16.4}h_1^{(1)}(z,\widetilde{\zeta}) \stackrel{def}{=}
\widetilde{\zeta}^4 -z=0
\end{equation}
This equation has four solutions in the form
\begin{equation}
\label{16.5}\widetilde{\zeta}(z)\equiv \widetilde{g}_{1/4} z^{1/4}
\end{equation}
where $\widetilde{g}_{1/4}^4=1$, i.e.
$\widetilde{g}_{1/4}^{(l=\{1,2\})}=\pm1$,
$\widetilde{g}_{1/4}^{(l=\{3,4\})}=\pm i$.

The shifted carrier of equation \eqref{16.2} lies in the lattice
with the basis vectors $(1,1)$ and $(0,5)$. Together with vector
(-1/4,1), corresponding to the shifted carrier of reduced
solutions \eqref{16.5}, they generate the new lattice with the
basis vectors $(-1/4,1)$ and $(0,5)$. The cone of the problem here
is $\mathcal{K}=\left\{k<\frac14\right\}$, then we get
\begin{equation}
\label{16.6}\mathbf{K}=\left\{\frac{1-5k}4,k\in\mathbb{N}\right\}
\end{equation}
Hence power expansions, conforming to reduced solutions
\eqref{16.5} can be written as
\begin{equation}
\label{16.7}\widetilde{\zeta}(z)\equiv\widetilde{\zeta}^{(l)}(z)=\widetilde{g}^{(l)}_{1/4}
z^{1/4}
+\sum_{k\,\in\,\,\mathbb{N}}\widetilde{g}^{(l)}_{(1-5k)/4}\,z^{(1-5k)/4},\qquad
 l=1,2,3,4
\end{equation}
where coefficients $\widetilde{g}^{(l)}_{(1-5k)/4}$, $k\in
\mathbb{N}$ can be uniquely determined.

 Taking into account three
terms, we can write
$$\widetilde{\zeta}^{(l)}(z)= \widetilde{g}^{(l)}_{1/4}
z^{1/4}-\frac{3}{8}z^{-1}+{\frac {5}{128}}\,{\frac
{32\,{\alpha}^{2}-32\,\alpha-9}{\widetilde{g}^{(l)}_{1/4}}}
z^{-9/4}+\ldots$$

As the process of calculations of the exponential additions of the
second and third levels coincides with that described in sections
10 and 11, let us just list the results here.

The second level of additions takes the form
\begin{equation}
\begin{gathered}
\widetilde{\xi}(z)\equiv\widetilde{\xi}^{(l,\,m)}(z)=\widetilde{r}_{1/4}^{(l,\,m)}
z^{1/4}+\frac{1}{4} z^{-1}
+\sum_{k\,\in\,\,\mathbb{N}}\widetilde{r}^{(l,\,m)}_{-1-5k/4}\,z^{-1-5k/4},\\
\qquad l=1,2,3,4;\,\,\quad\, m=1,2,3
\end{gathered}
\end{equation}
where coefficient $\widetilde{r}_{1/4}^{(l,\,m)}$ can take on
three possible meanings
$$ \widetilde{r}_{1/4}^{(l,\,m=1)}=-2 \widetilde{g}_{1/4}^{(l)},\qquad
\widetilde{r}_{1/4}^{(l,\,m=2,3)}=(-1\pm i)
\widetilde{g}_{1/4}^{(l)}
$$ and coefficients $\widetilde{r}^{(n,\,l,\,m)}_{-1-5k/4}$,
$k\in \mathbb{N}$ are uniquely determined.

The exponential additions of the third level can be presented as
\begin{equation}
\begin{gathered}
\widetilde{\eta}(z)\equiv
\widetilde{\eta}^{(l,\,m,\,p)}(z)=\widetilde{q}^{(l,\,m,\,p)}_{1/4}
z^{1/4}+\frac{1}{4}z^{-1}
+\sum_{k\,\in\,\,\mathbb{N}}\widetilde{q}^{(l,\,m,\,p)}_{-1-5k/4}\,z^{-1-5k/4},\\
n,l=1,2,3,4;\,\, m=1,2,3;\,\, p=1,2
\end{gathered}
\end{equation}
where
$$ \widetilde{q}^{(l,\,m,\,p={1,2})}_{1/4}= -\frac32\,\widetilde{r}_{1/4}^{(l,\,m)}-2\,\widetilde{g}_{1/4}^{(l)}
\pm\,\frac12i\sqrt{{3\,(\widetilde{r}_{1/4}^{(l,\,m)})^{2}+8\,{\widetilde{r}_{1/4}^{(l,\,m)}}\,\widetilde{g}_{1/4}^{(l)}+8\,
(\widetilde{g}_{1/4}^{(l)})^{2}}}$$ and coefficients
$\widetilde{q}^{(l,\,m,\,p)}_{-1-5k/4}$, $k\in \mathbb{N}$ can be
uniquely computed.

Taking into account three levels of exponential additions, we can
write solutions $w(z)$ at $z\rightarrow\infty$ in the form
\begin{multline*}
\label{}w(z)=\widetilde{\varphi}(z)+\\+\exp\left[ \int dz \left(
\widetilde{\zeta}^{(l)}(z)+ \exp \left[\int
dz\left(\widetilde{\xi}^{(l,\,m)}(z)+\exp\left(\int
dz\,\widetilde{\eta}^{(l,\,m,\,p)}(z)\right)\right)\right]\right)\right]
\end{multline*}
where $l=1,2,3,4;\,\,\,m=1,2,3;\,\,\,p=1,2$.

\section{Expansions to solutions of equation \eqref{e:eq} near arbitrary point $z=z_0$.}
In sections 3--9 we obtain all the expansions near points $z=0$
and $z=\infty$. To obtain the expansions to solutions of equation
\eqref{e:eq} near arbitrary point $z=z_0$, we use the substitution
$z'=z-z_0$. Then we get equation
\begin{equation}
\label{e:eq1}w_{z'z'z'z'} - 5w^2w_{z'z'} +5 w_{z'}w_{z'z'}-
5ww^2_{z'} +  w^5  - z'w-z_0w - \alpha =0
\end{equation}
Comparing this equation with equation \eqref{e:eq} we see, that it
has the additional term $-z_0w$, therefore, the carrier of this
equation contains all the points of the carrier of equation
\eqref{e:eq} and one additional point $Q_7=(0,1)$. The convex hull
of equation \eqref{e:eq1} coincides with quadrangle, presented at
fig. \ref{fig:ff1}, if $\alpha\ne0$, and with triangle, given on
fig. \ref{fig:ff3}, if $\alpha=0$. Point $Q_7=(0,1)$ is the inner
point, so it has no influence on asymptotic forms of solutions of
equation \eqref{e:eq1}.

If $z\rightarrow z_0$, then $z'\rightarrow 0$. Analyzing the
expansions to solutions of equation \eqref{e:eq}, we see, that
expansions near $z'=z-z_0=0$ correspond to apex $Q_1$ and edges
$\Gamma_1$ and $\Gamma_4$ only.

As described above,  there are four asymptotic forms,
corresponding to apex $Q_1$ (see section 3) and one asymptotic
form, corresponding to edge $\Gamma_4$ (see section 7). Power
expansions, conforming to these asymptotic forms, can be generally
written as
\begin{equation*}\begin{gathered}
w(z)=c_{{0}}+c_{{1}}(z-z_0)+c_{{2}}{(z-z_0)}^{2}+c_{{3}}{(z-z_0)}^{3}+
\\+ \left( \alpha-10\,c_{{1}}c_{{2}}+z_0\,c_{{0}}-{c_{{0}}}
^{5}+10\,c_{{2}}{c_{{0}}}^{2}+5\,c_{{0}}{c_ {{1}}}^{2} \right)
\frac{{(z-z_0)}^{4}}{24}+\\
+ \left( z_0\,c_{{1}}
-20\,{c_{{2}}}^{2}+40\,c_{{2}}c_{{0}}c_{{1}}+30\,c_{{3}}{c_{{0}}}^{
2}+c_{{0}}-30\,c_{{1}}c_{{3}}+5\,{c_{{1}}}^{3}-5\,{c_{{0}}}^{4}c_{{1}}
\right) \frac{{(z-z_0)}^{5}}{120} +\ldots
\end{gathered}\end{equation*}
where $c_0$, $c_1$, $c_2$ and $c_3$ are the arbitrary constants.
If $c_1=c_1=c_2=c_3=0$, this expansion corresponds to edge
$\Gamma_4$, otherwise it conforms to apex $Q_1$.

Edge $\Gamma_1$ also generates four asymptotic forms, and, using
the discussion, similar to presented in section 4, we obtain four
families of expansions
\begin{multline*}w(z)=(z-z_0)^{-1}+c_{{1}}(z-z_0)+c_{{2}}{(z-z_0)}^{2}+\\+ \left(
\frac{{c_{{1}}}^{2}}{4}-\frac{z_0}{20}
 \right) {(z-z_0)}^{3}- \left( \frac{\alpha}{36}+\frac{1}{36} \right) {(z-z_0)}^{4}+c_{{5}}{(z-z_0)}
^{5}+\\+ \left( {\frac {1}{160}}\,z_0\,c_{{2}}-{\frac
{7}{2880}}\,c_ {{1}}-{\frac
{5}{576}}\,c_{{1}}\alpha+\frac{1}{32}{c_{{1}}}^{2}c_{{2}}
 \right) {(z-z_0)}^{6}
+\ldots
\end{multline*}
\begin{multline*}w(z)=-2(z-z_0)^{-1}+c_{{1}}(z-z_0)+c_{{2}}{(z-z_0)}^{2}+\\+ \left(
\frac{z_0}{10}-\frac{7{c_{{1}}}^{2}}{2}
 \right) {(z-z_0)}^{3}+ \left( \frac{1}{18}-\frac{5 c_{{1}}c_{{2}}}{2}-\frac{\alpha}{36}
 \right) {(z-z_0)}^{4}+\\+c_{{5}}{(z-z_0)}^{5}+ \left( {\frac {21}{4}}\,{c_{{1}}}^{2}
c_{{2}}-{\frac {7}{80}}\,z_0\,c_{{2}}-{\frac {71}{1440}}\,c_{{1}}
+\frac{1}{36}\,c_{{1}}\alpha \right) {(z-z_0)}^{6}+\ldots
\end{multline*}
\begin{multline*}w(z)=-3(z-z_0)^{-1}-\frac {z_0}{60}{(z-z_0)}^{3}+ \left( {\frac {\alpha}{84}}
 -\frac{1}{28}\right)
{(z-z_0)}^{4}+\\+c_{{5}}{(z-z_0)}^{5}+c_{{6}}{(z-z_0)}^{6}-{\frac
{z_0^2}{4800}} {(z-z_0)}^{7}+{\frac {z_0}{554400}}\left( 95
\,\alpha-299 \right) {(z-z_0)}^{8}+ \ldots
\end{multline*}
\begin{multline*}w(z)=4 (z-z_0)^{-1}+{\frac {z_0}{220}}{(z-z_0)}^{3}+ \left( {{\frac {\alpha
}{504}} +\frac {1}{126}}\right)
{(z-z_0)}^{4}+\\+c_{{5}}{(z-z_0)}^{5}+{\frac
{17z_0^2}{5227200}}{(z-z_0)}^{7}+{\frac {z_0}{2217600}}\left(
13+5\, \alpha \right) {(z-z_0)}^{8}+ \\+\left( {\frac
{137}{47500992}}\,\alpha+{\frac {1}{2992}}\,{z_0}\,c_{{5}}+{\frac
{37}{11875248}}+{\frac {25}{ 47500992}}\,{\alpha}^{2} \right)
{(z-z_0)}^{9}+\\+{\frac {c_{{5}}}{15120}}
 \left( 6+5\,\alpha \right) {(z-z_0)}^{10}+c_{{11}}{(z-z_0)}^{11}+
 \ldots
\end{multline*}
where $c_1$, $c_2$, $c_5$, $c_6$ and $c_{11}$ are the arbitrary
constants.

All the expansions, presented in this section, converge for
sufficiently small $|z|$.

\section{Conclusion.}

In this paper we have shown that the self-similar solutions of both the Savada-Kotera
equation and the Kaup-Kupershmidt equation can be expressed via the solutions of one of
the fourth-order analogues of the second Painlev\'{e} equation. By means of power
geometry methods we have got all power and non-power asymptotic forms of solutions of
studied equation and all power expansions, corresponding to these expansions.

Let us briefly review the obtained results.

Near $z=0$ we have found four-parametric family $G_1^{(0)}1$,
three-parametric families $G_1^{(0)}2$, $G_1^{(1)}1$,
$G_1^{(1)}2$, two-parametric families $G_1^{(0)}3$, $G_1^{(1)}3$,
$G_1^{(1)}4$, one-parametric family $G_1^{(0)}4$ and family
$G_4^{(1)}1$(if $\alpha\ne0$). Families $G_1^{(0)}2$,
$G_1^{(0)}3$, $G_1^{(0)}4$ and $G_4^{(1)}1$ are the special cases
of $G_1^{(0)}1$.

Similar expansions have been obtained near arbitrary point $z=z_0\ne\infty$.

Near point $z=\infty$ we have found families of expansions $G_2^{(1)}n$ $(n=1,2,3,4)$
and $G_3^{(1)}1$ (if $\alpha\ne0$). For each of these expansions three levels of
exponential additions have been computed. If $\alpha=0$, four families of non-power
asymptotic forms near $z=\infty$ have been calculated too.

Expansions $G_2^{(1)}n$ $(n=1,2,3,4)$ were found earlier
\cite{Kudryashov01,Kudryashov10}, while all other expansions are
the new ones.

The comparison between expansions, obtained in this paper, and
ones, corresponding to the Painlev\'{e} equations
\cite{Bruno04,Bruno05,Bruno06,Bruno07,Gromak01} and their other
fourth-order analogues \cite{Efimova01,Demina01}, confirms the
hypothesis that the fourth-order equation \eqref{e:eq} determines
new transcendental functions.

\bigskip

This work was supported by the International Science and
Technology Center (project B1213).

\label{lastpage}
\end{document}